\documentclass[11pt]{article}
\usepackage{graphicx}
\usepackage{amsmath}
\usepackage{amssymb}
\usepackage{rotating}
\usepackage{epsfig}
\usepackage{times}
\usepackage{euler}
\usepackage{eucal}
\usepackage{a4}
\newcommand{\Scri}{\mathcal{I}}
\begin{document}
\title{Corvino's construction using Brill waves}
\author{
{\normalsize Domenico Giulini}\\
{\normalsize Institute of Physics}\\
{\normalsize University of Freiburg, Germany}\\
{\normalsize giulini@physik.uni-freiburg.de}
\and
{\normalsize Gustav Holzegel}\\
{\normalsize Department of Applied Mathematics and Theoretical Physics}\\
{\normalsize University of Cambridge, England}\\
{\normalsize G.Holzegel@damtp.cam.ac.uk}
}

\maketitle

\begin{abstract}
\noindent
For two-black-hole time-symmetric initial data we consider the 
Corvino construction of gluing an exact Schwarzschild end. 
We propose to do this by using Brill waves. We address the 
question of whether this method can be used to reduce the overall 
energy, which seems to relate to the question of whether it 
can reduce the amount of `spurious' gravitational radiation.
We find a positve answer at first order in the inverse gluing 
radius. 

\end{abstract}

\section{Introduction}
\label{sec:Introduction}
Processes involving collisions and mergers of black holes are 
amongst the most promising candidates regarding the generation 
of sufficiently strong gravitational waves. Hence much analytical 
and numerical effort went into the modeling of such processes 
as Cauchy problems for the matter-free Einstein equations.  
Initial data consist of a triple $(\sigma,h,K)$, where 
$\Sigma$ is a three manifold (the Cauchy hypersurface) and $h,K$
are two symmetric second-rank tensor fields satisfying a system 
of four underdetermined-elliptic differential equations, the 
constraints. $h$ is a Riemannian metric for $\Sigma$ and $K$ 
essentially its time derivative. Here we shall only be concerned 
with data representing two black holes at the moment of 
time symmetry. The latter condition is equivalent to $K=0$. 
Physically this corresponds to the most simple situation where 
two uncharged and unspinning black holes are placed at some 
distance, without relative velocity and without any overall linear 
or angular momentum. We will even assume equal hole masses so that 
the only free parameters are the mass and the mutual distance. 

Data representing two holes at the moment of time symmetry 
have been known for long and can be written down 
analytically (see e.g. \cite{Giulini:2003} for a recent review). 
One might think that the really hard part is their time 
evolution by means of a sophisticated combination 
of analytical and numerical techniques. However, there are 
also unsolved conceptual and analytical problems regarding just 
the initial data. One of them is the problem of `spurious' 
gravitational radiation, which we address here. 

%% Figure Scri-plus
\begin{figure}[htb]
\begin{center}
\epsfig{figure=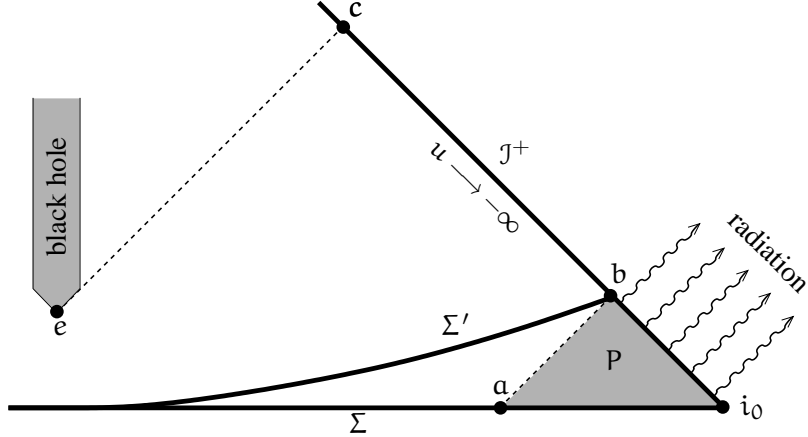,width=0.8\linewidth}
\put(-277,60){\begin{rotate}{90}black hole \end{rotate}}
\put(-27,73){\begin{rotate}{-45}radiation \end{rotate}}
\put(-140,100){\small\begin{rotate}{-45}$u\longrightarrow-\infty$\end{rotate}}
\put(-170,-7){$\Sigma$}
\put(-135,30){$\Sigma'$}
\put(-22,0){$i_0$}
\put(-71,50){$b$}
\put(-116,8){$a$}
\put(-170,151){$c$}
\put(-282,30){$e$}
\put(-170,151){$c$}
\put(-73,17){$P$}
\put(-113,95){$\Scri^+$}
\caption{
Conformal diagram of spacetime with two spacelike hypersurfaces 
$\Sigma$ and $\Sigma'$ ending at spacelike infinity $i_0$ and 
future lightlike infinity $\Scri^+$ respectively. The difference 
between the ADM-energy of $\Sigma$ (computed at $i_0$) and the 
Bondi mass of $\Sigma'$ (computed at the intersection 2-sphere 
between $\Sigma'$ and $\Scri^+$, here denoted by the point $b$) 
must be due to gravitational radiation escaping between $i_0$ 
and $b$. This radiation originates form the causal past $P$ 
of the region $\overline{i_0b}\subset\Scri^+$, whose intersection 
with $\Sigma$ is the region $\overline{ai_0}$. In this sense we 
say that the data on $\Sigma$ contained radiation in that region. 
Any gravitational radiation emerging from an (quasi localized) 
event $e$, e.g. the formation of a black hole due to neutron-star 
mergers, cannot register before $c$, the intersection of the 
future light-cone at $e$ with $\Scri^+$.}
\label{fig:Radiation}
\end{center}
\end{figure}

It has recently been shown to some level of rigour that even 
the most simple two black-hole collision data contain radiation 
in any neighbourhood of spacelike infinity $i^0$; 
see Figure\,\ref{fig:Radiation}. This was done by approximately 
evaluating the Bondi mass on a cut of $\Scri^+$ close to $i^0$
for time symmetric, conformally flat data. The result 
was~\cite{ValienteKroon:2003}
\begin{equation}
\label{eq:BondiADM}
m_{\text{Bondi}}=m_{\text{ADM}}+\sum_{k=-2}^{k=2}
\vert G_k\vert^2
\left(\frac{\sqrt{2}}{u}\right)^7+O(1/u^8)\,,
\end{equation}
where the $G_k$ are the Newman-Penrose constants
\cite{NewmanPenrose:1968} and where $u$ is the Bondi parameter along 
$\Scri^+$, which tends to $-\infty$ as one approaches $i^0$. 
Hence $m_{\text{Bondi}}\leq m_{\text{ADM}}$ with equality 
at $i^0$, as it must be. If some of the Newman-Penrose constants 
do not vanish, we have a strict inequality off $i^0$, meaning that some 
gravitational radiation has reached $\Scri^+$ in between. 
If this is the case, one says that the initial data contained 
radiation in a neighbourhood of (spacelike) infinity. 
This radiation is spurious in the sense that it cannot be caused 
by the collision process. Hence one would like to remove it. 
If we were in a linear theory, e.g. electrodynamics, the meaning 
of `remove' would be to subtract the appropriate solution to the 
homogeneous equations. However, in a non-linear theory, like 
General Relativity, it is a\,priori unclear how to analytically 
achieve a `removal'. Is it possible at all to get rid of the 
spurious radiation without touching the interior data? 

The question addressed here relates to the problem of analytically 
understanding the region where null infinity touches spacelike 
infinity. The first step in this direction was taken by Friedrich 
\cite{Friedrich:1998}, who used the conformal field equations to 
analyse this region. In this approach, spacelike infinity is blown 
up to a cylinder $S^2\times[0,1]$, whose two boundary
components connect to future and past null infinity ($\Scri^\pm$) 
respectively. The Cauchy surface, $\Sigma$, intersects the cylinder 
in a 2-sphere cross section, $i^0$, from which certain transport 
equations govern the propagation of initial data along the cylinder 
to its boundary. This propagation will generically lead to logarithmic 
divergences at these intersections and, consequently, to a non-smooth 
null infinity. Whether, and to what degree, a non-smooth null infinity 
is physically objectionable is currently unclear. 
Some necessary conditions for the avoidance of such singularities 
are known~\cite{ValienteKroon:2004a}, but so far no theorem establishes 
a sufficient set. Instead, the following conjecture was 
launched~\cite{ValienteKroon:2004a}: 

\vspace{0.3cm}
\noindent
\textbf{Conjecture:} 
If an initial data set which is time-symmetric and confomally flat 
in a neighbourhood $B_a(i_0)$ of the point $i_0$ yields a development 
with a smooth null infinity, then the initial data are, in fact, 
Schwarzschild in $B_a(i)$.
\par\vspace{0.3cm} 

If true, the conjecture tells us that gravitational radiation is 
excluded near spacelike infinity. 
At this point one might fear that the condition of being asymptotically 
Schwarzschild might exclude most data of physical interest. 
But fortunately this is not true, thanks to the work of Corvino 
\cite{Corvino:2000}, who proved that is possible to glue a 
Schwarzschild metric along an annulus to an asymptotically flat, 
conformally flat three metric if the gluing radius is chosen 
sufficiently large and if the mass of the Schwarzschild metric 
is chosen appropriately. That is, there do exist spacetimes which are 
Schwarzschild at infinity (therefore ensuring a smooth null-infinity) 
and non-static in the inside. 

For example, we could glue a Brill-Lindquist initial data set 
for the collision of two black holes to a Schwarzschild metric. 
This would clearly \emph{remove} the `spurious' radiation at spacelike 
infinity. However, it might at the same time \emph{introduce}
`spurious' radiation through the modified gluing metric on the 
annulus, which is generally very complicated. In particular, it is 
not conformally flat. This motivates to ask about the mass of the 
glued Schwarzschild solution, depending on the specific gluing 
instructions. 

Information about the mass is not easy to obtain: the gluing 
function is hard to control and not easily calculated explicitly. 
What we need is a physical idea of how the gluing function might 
look like. Here we propose to specialize to Brill waves. They are 
axisymmetric (as is Brill-Lindquist) and not conformally flat. 
Is it possible to construct a spacetime which is Brill-Lindquist 
to some radius, has Brill-form on an annulus, and is Schwarzschild 
outside? It is the purpose of this paper to explore this idea a 
little further, based on the diploma thesis by one of 
us~\cite{gustav}.

\section{Corvino's construction}
\label{sec:CorvinoConstruction}
We mentioned that the Brill-Lindquist data seem to contain spurious 
gravitational radiation near spacelike infinity and, intimately 
related to this fact, cannot lead to an asymptotically simple 
spacetime. Hence we wish to modify them near infinity. The techniques 
of Corvino use the \emph{underdetermined} ellipticity of the constraint 
equations to glue a Schwarzschild end to an arbitrary asymptotically 
flat manifold. 

Consider a smooth gluing function $\beta \geq 0$, $|\beta|\leq 1$,
where
\begin{equation}
\label{eq:BumpFunction}
\beta = 
\left\{ 
\begin{array}{ll}
1 & \textrm{for $x \leq 1$} \\
0 & \textrm{for $x \geq 2$}
\end{array} 
\right.
\end{equation}
and an arbitrary asymptotically flat metric $g_{AF}$ together with a 
Schwarzschild metric $g_{SS}^{m,\vec{c}}$ with ADM-energy $m$ and center 
$\vec{c}$. Then, in an asymptotically flat chart, the metric
\begin{equation}
\label{eq:GlueMetric}
g_{glue}=
\beta\left(\frac{|x|}{R}\right)g_{AF} + 
\left(1-\beta\left(\frac{|x|}{R}\right)\right) g_{SS}^{m,\vec{c}}
\end{equation}
(for constant $R$) is smooth and glues $g_{AF}$ to $g_{SS}^{m,\vec{c}}$. 
However, the constraint equation of vanishing scalar curvature, 
$R(g_{glue})=0$, is now violated on the annulus, $R<|x|<2R$, hence 
$g_{glue}$ is not a time-symmetric initial datum for Einstein's 
equation. Corvino now shows that it is possible to choose the gluing 
radius $R$ large enough to find a metric $h$, having support only 
on the annulus, such that $R(g_{glue}+h)=0$ everywhere. To make 
this construction work, the variables $m$ and $\vec{c}$ have to be 
chosen appropriately. In this way Corvino's construction provides 
us with time-symmetric initial data, isometric to Schwarzschild 
outside some finite radius and non-trivial in the inside. But his 
construction is not very explicit. We have no method at hand to 
construct the metric explicitly, because we do not know much more 
about the perturbation $h$ on the annulus than its existence.
This leaves important physical questions unanswered: Can we give a 
physical interpretation for the gluing construction? One is 
tempted to ask the following `physical' questions: does the 
construction actually remove spurious gravitational-radiation energy, 
or does it at best just lead to a redistribution of energy from 
infinity to the annulus region\,? The answer does not seem to be 
known. 

\section{Brill waves and the gluing construction}
\label{sec:BrillWavesGlueing}
\subsection{Brill-Lindquist data}
\label{sec:BrillLindData}
We propose to phrase these questions in a more explicit environment. 
Consider Brill-Lindquist data describing the head-on collision of 
two black holes: 
\begin{equation}
\label{eq:BL-Data}
g_{ij}=\left(1+\frac{a_1}{|\vec{x}-\vec{c}_1|}+
\frac{a_2}{|\vec{x}-\vec{c}_2|}\right)^4 \delta_{ij}\,.
\end{equation}
We further simplify by choosing $a_1=a_2$, corresponding to equal 
masses for the holes and $\vec c_1=-\vec c_2=\vec c:=(0,0,d/2)$, so 
that the holes lie symmetrically about the origin on the $z$-axis: 
\begin{equation}
g_{ij}=
\left(1+\frac{a}{|\vec{x}-\vec{c}|}
+\frac{a}{|\vec{x}+\vec{c}|}\right)^4 \delta_{ij}\,.
\end{equation}
These data are axisymmetric. We shall equip them with a 
Schwarzschild end at infinity.

\subsection{Brill waves}
\label{sec:BrillWaves}
Our key assumption is that the metric on the annulus is a conformally 
transformed Brill wave~\cite{Brill:1959}. In cylindrical coordinates it 
correspond to the regular 3-metric 
\begin{equation} 
\label{Bmet}
g_{Brill}=
e^{2 q(\rho, z)}\left(d\rho^2+dz^2\right)+\rho^2d\theta^2\,.
\end{equation}
Here $q$ is a function of $\rho$ and $z$ satisfying the following conditions.
\begin{alignat}{3} 
& q\,=\,0\quad 
&&\text{if}\quad \rho=0\quad 
&&\text{(on the z-axis)}\,,\nonumber\\
\label{brillcon}
& \frac{\partial}{\partial \rho}q\,=\,0\quad
&&\text{if}\quad \rho=0\quad
&&\text{(on the z-axis)}\,,\\
& q\rightarrow 0\ 
&&\hspace{-6mm}\text{faster than}\ \tfrac{1}{r}\quad 
&&\text{for}\ r \rightarrow \infty\,.\nonumber
\end{alignat}
The second condition is immediately obvious: If q is to represent 
an axisymmetric solution, it must have an extremum on the axis in 
the $\rho$ direction. The first condition ensures that there is no 
conical singularity on the z-axis. The third condition will be 
satisfied automatically because our $q$ will have compact support.

To solve the constraint equation $R=0$, we need to perform a conformal 
transformation of the Brill metric. We ask if there exists a conformal 
factor, $\psi$, with the following properties
\begin{equation}
\begin{split}
& R(\psi^4g_{Brill})=0\,,\\
& \psi > 0\quad\text{everywhere}\,,\\
& \psi \longrightarrow 1\quad\text{at}\ \,\infty\,.\\
\end{split}
\end{equation}
Using the behaviour of the Ricci scalar under conformal transformations, 
$R(\psi^4g)=-8\psi^{-5}\left(\bigtriangleup_g-\frac{1}{8}R(g)\psi\right)$, 
we end up with the equation
\begin{equation}
\label{eq:RicciTrans}
8\bigtriangleup_g \psi - R(g) \psi = 0 \,,
\end{equation}
where $\psi >0$ everywhere and $\psi\rightarrow 1$ at $\infty$. 
In the past much work has been done to derive existence theorems 
for solutions $\psi$ given $q$. Here we will take a slightly 
different viewpoint.

\subsection{Our construction}
\label{OurConstruction}
Combining the spirit of Corvino's construction with our physical 
idea of a Brill wave on the annulus, we define a conformal factor
\begin{equation} 
\label{conff}
\psi=\left(1+
\frac{a}{\left|\vec{x}-\vec{c}\right|}+
\frac{a}{\left|\vec{x}+\vec{c}\right|}\right)\beta
\left(\frac{\left|x\right|}{R}, \theta\right)+
\left(1-\beta\left(\frac{\left|x\right|}{R}, 
\theta\right)\right)\left(1+\frac{A}{\left|\vec{x}\right|}\right)\,.
\end{equation}
We have generalized the gluing function $\beta$ to be axisymmetric 
on the annulus:
\begin{equation}
\beta = \left\{ \begin{array}{ll}
1 & \textrm{for $x \leq 1$ and all $\theta$} \\
0 & \textrm{for $x \geq 2$ and all $\theta$}
\end{array} \right.
\end{equation}
and $\beta^{(n)}(1)=\beta^{(n)}(2)=0$ for all $n \geq 1$. 

Next we use the fourth power of the function $\psi$ in (\ref{conff})
as conformal factor to define the physical initial-data metric:
\begin{equation} 
\label{brillcompl}
g=\psi^4\,g_{\text{Brill}}
 =\psi^4\,\left(
e^{2q(\rho, z)}\left(d\rho^2+dz^2\right)+\rho^2d\theta^2\right)\,.
\end{equation}
Here $q$ is, by definition, a function with support on the annulus, vanishing 
derivatives on its boundary\footnote{In the sequel a function with these 
properties will sometimes be called a ``function of bump-character'' 
(although it can become negative, of course).} and which satisfies 
the Brill-conditions (\ref{brillcon}). Taking all this into account 
we observe that 
\begin{itemize}
\item $g$ is Brill-Lindquist for $r < R$,
\item $g$ is Schwarzschild for $r > 2R$ (with ADM-energy $M=\frac{A}{2}$),
\item $g$ is a complicated, conformally transformed Brill wave on 
      $r \in [R,2R]$.
\end{itemize}

Now, the constraint equation $R(g)=0$ is satisfied if
\begin{equation} 
\label{crit2}
\left(\frac{\partial^2}{\partial \rho^2}
+\frac{\partial^2}{\partial z^2}\right)q(\rho,z)
=\left(\frac{\partial^2}{\partial r^2}
+\frac{1}{r}\frac{\partial}{\partial r} 
+\frac{1}{r^2}\frac{\partial^2}{\partial \theta^2}\right)
q(r,\theta)=\frac{-\bigtriangleup^{(3)}\psi}{\psi}\,,
\end{equation}
Since $\psi$ stands for the expression (\ref{conff}), this equation 
should be read as relating the gluing function $\beta$ to the 
Brill-function $q$. It turns into an identity ($0=0$) for $r<R$ and 
$r>2R$. The mathematical structure of the equation is discussed in 
the next section. 

\subsection{The mathematical problem}
\label{sec;MathFormulation}
Equation (\ref{crit2}) can be understood as an elliptic equation on 
the two-dimensional plane. However, the boundary conditions are not of 
the usual Dirichlet or Neumann type. Note that the right hand 
side---for any gluing function $\beta$---is of bump-type, i.e. 
vanishes with all derivatives at $r=R$ and $r=2R$. Hence we have a 
large freedom in the choice of the right hand side by exploiting the 
freedom of the gluing function $\beta$, of which only the boundary 
values (including all derivatives) are fixed. On the other hand, there 
are boundary conditions on $q$ and its derivatives that have to be 
taken into account. 
\begin{figure}[htb]
\begin{center}
\epsfig{figure=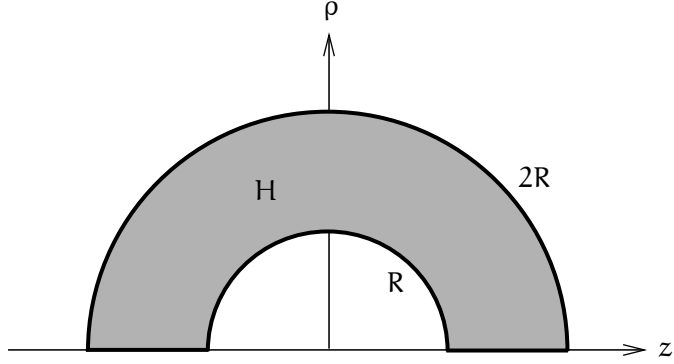,width=0.65\linewidth}    
\put(3,1){$z$}
\put(-124,130){$\rho$}
\put(-100,25){$R$}
\put(-50,65){$2R$}
\put(-150,60){$H$}
\caption{Half-annular region $H$ (shaded region) in upper 
$z$-$\rho$ halfplane.}
\label{fig:Annulus}
\end{center}
\end{figure}

As it turns out, the problem can be formulated in the following way: 
Given the Poisson equation 
\begin{equation} 
\label{v1}
\bigtriangleup^{(2)}q =f=-\frac{\bigtriangleup^{(3)} \psi}{\psi}
\end{equation}
in the half-annular region H sketched in Figure\,\ref{fig:Annulus}. 
Is it possible to choose the right hand side of (\ref{v1}) such that 
$q$ simultaneously satisfies Dirichlet and Neumann conditions for 
the given region H? This we will call a DN-solution. Note that imposing 
Dirichlet and Neumann conditions guarantees the Brill-conditions 
(\ref{brillcon}) on the z-axis, as well as the vanishing of all 
radial derivatives on the arches of H. Indeed, if $q$ and its first 
derivative vanish on the arches, then, by equation (\ref{v1}), all 
derivatives of $q$ will vanish there since $f$ vanishes with all 
derivatives on the arches. Although one can easily write down some 
$f$'s admitting a DN-solution, it is not clear if these $f$ can be 
obtained by $-\frac{\bigtriangleup^{(3)} \psi}{\psi}$ with an appropriate 
choice of $\beta$---the root of the problem being that the 
inhomogeneity is highly non-linear in $\beta$.

\subsection{Approximate results}
\label{sec:ApproxResults}
Corvino's construction only works for large gluing radii $R$. We could 
therefore try to solve the DN-problem to first order in the inverse
gluing radius. To do this we expand the inhomogeneity to first order 
in $1/R$, making two further choices:  

\begin{itemize}
\item[1]
The first is related to the possibility of reducing the energy. 
The further outside the gluing is performed, the smaller the 
difference of the Schwarzschild and the Brill-Lindquist data 
will be. In other words, the difference will scale in some way 
with the inverse gluing radius---the exact form being dependent 
on the details of the gluing procedure. Recall equation (\ref{conff}), 
introduce the scaling $|x| \rightarrow |x|R$ and 
expand in the inverse gluing radius, $\frac{1}{R}$:
\begin{equation}
\psi = 1 + \beta(r,
\theta)\left(\frac{2a-A}{Rr}+\frac{ad^2}{2R^3r^3}
+\mathcal{O}\left(\frac{1}{R^5}\right)\right)
\end{equation}
Note that we have no $\frac{1}{R^2}$-(dipole)-terms due to the choice
of equal masses for the two Brill-Lindquist holes and equal 
mass-centres for the Brill-Lindquist system and the Schwarzschild 
hole. We choose the scaling to be
\begin{equation}
2a-A=\frac{b}{R^2} 
\end{equation}
where $b$ takes real values in a small interval around $0$
(eventually we will be interested in $b>0$).\footnote{If we chose 
$2a-A=\frac{b}{R}$ we would, at the lowest order, have the same 
situation as for the gluing of two Schwarzschild black holes of 
ADM-energies $2a$ and $A$ respectively. On account of the Penrose
inequality one expects this not to work with mass reduction.
 Note that for large gluing radii the conformally transformed 
Brill wave will certainly not introduce a new apparent 
horizon. In fact, in can be shown that the recursion technique 
presented below fails for the Schwarzschild-Schwarzschild case.} 

\item[2]
The second choice is an ansatz for the gluing function 
$\beta$:
\begin{equation}
\beta(r,\theta)=\alpha(r)+\mu(r)\sin^2 \theta\,.
\end{equation}
Here $\alpha(r)$ is a radial gluing function ($\alpha(1)=1$, 
$\alpha(2)=0$, all derivatives vanish at these two points) and 
$\mu(r)$ is a function with bump character on the annulus 
($\mu(1)=0$, $\mu(2)=0$, all derivatives vanish at these two points) 
but otherwise arbitrary. For the expansion (\ref{conff}) of $\psi$ 
we explicitly introduce the gluing radius $R$ by scaling 
$r \rightarrow Rr$ in the formulae. This makes the gluing occur 
on the interval $\left[1,2\right]$).
\end{itemize}

The right hand side can then be written as
\begin{alignat}{1} 
\label{Case1}
\frac{\bigtriangleup\psi}{\psi}
=\frac{1}{R^5}&\left(\alpha''(r)+\frac{2}{r}\alpha'(r)
+\frac{4}{r^2}\mu(r)\right)\left(\frac{b}{r}
+\frac{ad^2}{2r^3}\right)
+2\alpha^{\prime}(r)\left(-\frac{b}{r^2}
-\frac{3ad^2}{2r^4}\right)
\nonumber \\
+\sin^2\theta &\Bigg[\left(\mu^{\prime\prime}(r)
+\frac{2}{r}\mu^{\prime}(r)-\frac{6}{r^2}\mu(r)\right)\left(\frac{b}{r}
+\frac{ad^2}{2r^3}\right)-\frac{3ad^2}{4r^3}
\left(\alpha^{\prime \prime}(r)+\frac{2}{r}\alpha^{\prime}(r)\right)
\nonumber \\
&\qquad +2\mu^{\prime}(r)
\left(-\frac{b}{r^2}-\frac{3ad^2}{2r^4}\right)
+\frac{9ad^2\alpha^{\prime}(r)}{2r^4}-\frac{9ad^2 \mu(r)}{r^5}\Bigg]
\nonumber \\
+\sin^4 \theta &\Bigg[-\frac{3ad^2}{4r^3}
\left(\mu^{\prime\prime}(r)+\frac{2}{r}\mu^{\prime}(r)
-\frac{6}{r^2}\mu(r)\right)+\frac{9ad^2\mu^{\prime}(r)}{2r^4}
+\frac{6ad^2}{r^5} \mu(r) \Bigg]
\nonumber \\
&\qquad+\mathcal{O}(R^{-6})
\end{alignat}
the right-hand side of which may be written in short-hand: 
\begin{equation} 
\label{shrhs}
\begin{split}
\frac{1}{R^5}\Bigl(
 A_0(\alpha',\alpha'',\mu)
&+A_2(\alpha',\alpha'',\mu,\mu',\mu'')\sin^2\theta\\
&+A_4(\mu, \mu',\mu'')\sin^4\theta\Bigr)
+\mathcal{O}(R^{-6})\,.
\end{split}
\end{equation}

We make the following ansatz for $q(r,\theta)$:
\begin{equation}
q(r, \theta)=\frac{1}{R^3}\left(B_1(r/R)\sin^2 
\theta+B_2(r/R)\sin^4 \theta \right)
\end{equation}
The functions $B_i$ are of bump-character in $\left[1,2\right]$. 
This ansatz satisfies the three Brill-conditions (\ref{brillcon}) 
for $q(r,\theta)$. The expression will be of order $\frac{1}{R^5}$ 
when acted upon by the Laplacian. If we apply the Laplacian and 
perform the scaling with the gluing radius afterwards, we obtain 
the following left hand side:
\begin{align}
LHS= \frac{1}{R^5}\Bigg[\frac{2B_1(r)}{r^2}\,+\,&\sin^2 \theta 
\left(B_1^{\prime \prime}(r)+\frac{2}{r}B_1^{\prime}(r)
-\frac{4}{r^2}B_1(r)+\frac{12B_2(r)}{r^2}\right)
\nonumber \\
+\,&\sin^4 \theta\left(B_2^{\prime\prime}(r)+\frac{1}{r}B_2^{\prime}(r)-
16\frac{B_2(r)}{r^2}\right)\Bigg]\,.
\end{align}
The right-hand side is given by minus the expansion (\ref{shrhs}) with 
\begin{alignat}{2}
& A_0&&\,=\,
\left(\alpha^{\prime \prime}(r)+\frac{4}{r^2}\mu(r)\right)
\left(\frac{b}{r}+\frac{ad^2}{2r^3}\right)-2\alpha'(r)
\frac{ad^2}{r^4} \\
& A_2&&\,=\,\left(\mu^{\prime\prime}(r)-\frac{6}{r^2}\mu(r)\right)
\left(\frac{b}{r}+\frac{ad^2}{2r^3}\right)\nonumber\\
& &&\qquad -\frac{3ad^2}{4r^3}
\left(\alpha^{\prime \prime}(r)-\frac{4}{r}\alpha^{\prime}(r)
+\frac{12}{r^2} \mu(r)\right)-2\mu^{\prime}(r)\frac{ad^2}{r^4} \\
& A_4&&\,=\,-\frac{3ad^2}{4r^3}\left(\mu^{\prime\prime}(r)+\frac{2}{r}
\mu^{\prime}(r)-\frac{6}{r^2}\mu(r)\right)
+\frac{9ad^2\mu^{\prime}(r)}{2r^4}+\frac{6ad^2}{r^5} \mu(r)
\end{alignat}
The Dirichlet-Neumann problem for this part of the inhomogeneity is 
solved if the following differential equations hold 
(Note the additional minus sign since the right hand side is 
$-\frac{\bigtriangleup \psi}{\psi}$):
\begin{alignat}{2}
& \frac{2}{r^2}B_2(r)&&\,=\,-A_0(\alpha',\alpha'',\mu)\\
& B_2''(r)+\frac{1}{r}B_2^{\prime}(r)
-\frac{4}{r^2}B_2(r)+\frac{12B_4(r)}{r^2}
&&\,=\,-A_2(\alpha',\alpha'',\mu,\mu',\mu'')\\
& B_4''(r)+\frac{1}{r}B_4'(r)-16\frac{B_4(r)}{r^2}
&&\,=\,-A_4(\mu, \mu',\mu'')
\end{alignat}
We interpret the first two equations as defining equations for 
$B_2$ and $B_4$ given any $\alpha$ and $\mu$. The functions 
$\alpha$ and $\mu$ are then determined by the third differential 
equation: We relate $\alpha$ and $\mu$ in such a way that the 
third differential equation holds. In this case $q$ (constructed 
from $B_2$, $B_4$) is a solution of the Dirichlet-Neumann problem. 

Carrying out the calculation explicitly we arrive at
\begin{equation} 
\label{iode}
\begin{split}
& \frac{ad^2+2br^2}{24r} \mu^{(4)}(r) 
+ \frac{2br^2-7ad^2}{24r^2}\mu^{(3)}(r)
+\frac{55ad^2-34br^2}{24r^3}\mu''(r)\\
&-\frac{85ad^2-32br^2}{12r^4}\mu'(r)
-\frac{245ad^2}{12r^5}\mu(r)\\
=\quad 
&-\frac{1}{48}r\left(ad^2+2br^2\right)\alpha^{(6)}(r)
-\frac{1}{24}\left(-ad^2+10br^2\right)\alpha^{(5)}(r)\\
&+\left(\frac{ad^2}{6r}-\frac{5br}{24}\right)\alpha^{(4)}(r)
+\left(\frac{15b}{8}
-\frac{7ad^2}{6r^2}\right)\alpha^{(3)}(r)\\
&+\frac{3\left(8ad^2-5br^2\right)}{8r^3}\alpha^{(2)}(r)
-\frac{3ad^2}{r^4}\alpha^\prime(r).\\
\end{split}
\end{equation}
What we have obtained is a differential equation relating the 
functions $\alpha$ and $\mu$. Can we find a gluing function $\alpha$ 
and a function $\mu$ of bump-character related by (\ref{iode})? 
We interpret this equation as an inhomogeneous ordinary differential 
equation for $\mu$. The right hand side---independent of the 
choice of the gluing function---will be a function of bump-character. 
If we set initial conditions $\mu^{(n)}(1)=0$ for $n=0,1,2,3$ all 
derivatives of $\mu$ will vanish at the point $r=1$ by the properties 
of the right hand side. The only additional condition $\mu$ has to 
satisfy is that the function and all its derivatives should also 
vanish at the point $r=2$. Can we choose a right hand side, such 
that $\mu$ and all its derivatives vanish also at $r=2$? 
The answer is provided by the solution formula for inhomogeneous 
ordinary differential equations. We multiply (\ref{iode}) by 
$\frac{24r}{ad^2+2br^2}\neq 0$ and obtain an equation of the 
following type
\begin{equation}
\mu^{(4)}(r)
+f(r)\mu^{(3)}(r)
+g(r)\mu^{(2)}(r)
+h(r)\mu^\prime(r)
+k(r)\mu(r)
=\mathcal{F}[\alpha](r)
\end{equation}
Here $f(r),g(r),h(r),k(r)$ are well behaved, bounded, rational
functions in the interval $[1,2]$. They also depend on the parameter 
$b$, which for notational simplicity we suppress to write out 
explicity most of the time. We will restore the argument $b$ when 
it becomes important, as it will below. (The functions also depend 
on the other parameters $a,d$. But this dependence does not interest 
us anyway.)  We recall that $\mathcal{F}[\alpha](r)$ 
is defined by the right hand side of (\ref{iode}) multiplied by 
$\frac{24r}{ad^2+2br^2}\neq 0$ and hence depends on the first to 
sixth derivatives of $\alpha$.\footnote{We exclude values 
$b\leq -ad^2/8$.} All derivatives of $\mathcal{F}[\alpha](r)$ with 
respect to $r$ vanish at the boundary $r=1,2$. The dependence of 
$\mathcal{F}$ on $b$ is also notationally suppressed at this moment.  

We transform the system to a system of four ordinary differential 
equations of first order in the usual way:
\begin{equation}
\label{eq:InhGlMatrix}
\begin{pmatrix}
\mu(r) \\
u(r)   \\
v(r)   \\
w(r) 
\end{pmatrix}^\prime =
\begin{pmatrix}
0 & 1 & 0 & 0 \\
0 & 0 & 1 & 0 \\
0 & 0 & 0 & 1 \\
-k(r) & -h(r) & -g(r) & -f(r)
\end{pmatrix}
\begin{pmatrix}
\mu(r) \\
u(r)   \\
v(r)   \\
w(r) 
\end{pmatrix}
+
\begin{pmatrix}
0 \\
0 \\
0 \\
\mathcal{F}[\alpha](r) 
\end{pmatrix}
\end{equation}
This we will write as 
\begin{equation}
\label{eq:InhGlVect}
\vec{x}^\prime(r)=A(r)\vec{x}(r)+
\vec{n}\left(\alpha(r)\right)\,.
\end{equation}
with the obvious identifications. Derivatives are always taken with
respect to $r$. We impose the boundary condition $\vec{x}(1)=0$. \\

The following standard arguments about ODEs are taken from 
\cite{Coddington:1955}. If $\Phi(r)$ is a fundamental 
matrix\footnote{A matrix, whose $n$ columns are the $n$ linearly 
independent solutions of the homogeneous system.} of the homogeneous 
system, then
\begin{equation} 
\label{solode}
\vec{x}(r)=
\Phi(r)\int_1^r \Phi^{-1}(t)\vec{n}\left(\alpha(t)\right)\,dt
\end{equation}
is a solution of the inhomogeneous system satisfying the boundary
condition $\vec{x}(1)=0$ imposed above. It then automatically follows 
from (\ref{solode}) that $\vec x$ has altogether vanishing derivatives 
at $r=1$. We also need to impose the condition that $\vec x$ and all 
its derivatives vanishes at the other boundary at $r=2$. How this can 
be achieved will be discussed next. But before we shall make 
the following observation: A priori our equations were real, so that 
$\Phi$ is a priori a real $4\times 4$ matrix. However, it will 
turn out to be useful to complexify our system. Note that even 
if we use complex fundamental matrices we will get a real solution 
$\vec{x}$ of (\ref{solode}). This follows from the fact that if 
$\Phi$ is some (complex) fundamental matrix, any other fundamental 
matrix is given by $\Phi C$ with some appropriate constant 
complex matrix $C$ (compare Thm.\,2.3 of \cite{Coddington:1955}). 
Obviously (\ref{solode}) is invariant under $\Phi\mapsto\Phi C$. 

Now we turn to imposing the second boundary condition: $\vec x(2)=0$.
Again we observe that this boundary condition together with 
(\ref{solode}) implies that all derivatives, too, vanish at $r=2$.
Requiring two boundary conditions for an ODE of first order is 
clearly an overdetermination. However, in our case, the 
inhomogeneity in (\ref{eq:InhGlVect}) is not fixed. Hence we read 
$\vec x(r=2)=0$ as equation restricting the choice of the function 
$\alpha$. If we define $\vec v$ to be the vector whose components 
are the fourth column of the matrix $\Phi^{-1}$, i.e. 
$v_i:=\Phi^{-1}_{i4}$, then the boundary condition is equivalent 
to 
\begin{equation}
\label{phic}
\int_1^2 \vec v(t)\mathcal{F}[\alpha](t)\,dt =\vec 0\,.
\end{equation}
Before we continue we remark that under a redefinition 
$\Phi\mapsto\Phi C$ we have $\Phi^{-1}\mapsto C^{-1}\Phi^{-1}$ 
and hence $\vec v\mapsto C^{-1}\vec v$, showing that (\ref{phic})
is satisfied for one choice of $\Phi$ if and only if it is 
satisfied for any other.  

To get an idea what may go wrong in trying to satisfy (\ref{phic}),
we use the explicit expression for $\mathcal{F}$ and get, after 
some integrations by parts:  
\begin{equation}
\label{eq:phicExpl}
\begin{split}
\int_1^2dr\ \alpha^\prime(r)\Bigg[
& \left(\frac{1}{2}r^2 \vec v(b,r)\right)^{(5)} 
+ \left(r\frac{ad^2-10br^2}{ad^2 + 2br^2}\vec v(b,r)\right)^{(4)}\\
-&\left(\frac{4ad^2-5br^2}{ad^2+2br^2} \vec v(b,r)\right)^{(3)} 
+\left(\frac{1}{r}\frac{35br^2-28ad^2}{ad^2+2br^2}\vec v(b,r) \right)^{(2)}\\
-&\left(\frac{9}{r^2}\frac{8ad^2-5br^2}{ad^2+2br^2}\vec v(b,r)\right)^{\prime}
-\frac{72}{r^3} \frac{ad^2}{ad^2+2br^2} \Bigg]=\vec 0\,.\\
\end{split}
\end{equation}
This clearly cannot be satisfied if one of the components functions 
in the square bracket is constant, since  
\begin{equation}
\label{eq:Contra}
\int_1^2 \alpha^\prime(r)dr = \alpha(2)-\alpha(1)= 1\,.
\end{equation}
More general, (\ref{eq:phicExpl}) cannot be satisfied if after left 
multiplication with some $C^{-1}$ at least one of the component 
functions is constant. However, in the sequel we shall prove that 
at least for small parameters $b$ this cannot happen.  The idea is 
to prove this directly for $b=0$ and then use an inverse-function type 
argument to extend this to small $b$. Hence we first need to 
consider the case $b=0$ in some detail.

\subsection{The case $b=0$}
\label{sec:CasebZero}
In the case $b=0$ the system (\ref{iode}) reduces to
\begin{equation}
\label{eq:bzero}
\begin{split}
&\mu^{(4)}(r)-\frac{7}{r}\mu^{(3)}(r)+\frac{55}{r^2}\mu^{\prime
  \prime}(r)-\frac{170}{r^3}\mu^\prime(r)-\frac{490}{r^4}\mu(r) 
=\\
&-\frac{1}{2}r^2\alpha^{(6)}(r)+r\alpha^{(5)}(r)+4\alpha^{(4)}(r)
-\frac{28}{r}\alpha^{(3)}(r)+\frac{72}{r^2}\alpha^{(2)}(r)
-\frac{72}{r^3}\alpha^{(\prime)}(r)\,.\\
\end{split}
\end{equation}
The homogeneous part is an equation of Euler-type, i.e. solved  
by simple powers $r^{n_i}$, so that the fundamental matrix is 
readily computed:
\begin{equation} 
\label{fuma}
\Phi(r)=
\begin{pmatrix}
r^{n_1} & r^{n_2} & r^{n_3} & r^{n_4} \\
(r^{n_1})' & (r^{n_2})' & (r^{n_3})' & (r^{n_4})'\\
(r^{n_1})'' & (r^{n_2})'' & (r^{n_3})'' & (r^{n_4})''\\
(r^{n_1})''' & (r^{n_2})''' & (r^{n_3})''' & (r^{n_4})'''
\end{pmatrix}
\end{equation}
%
%\Phi(r)=
%\begin{pmatrix}
%r^{n_1} & r^{n_2} & r^{n_3} & r^{n_4} \\
%n_1r^{n_1-1} & n_2r^{n_2-1} & n_3r^{n_3-1} &  n_4r^{n_4-1}\\
%n_1(n_1-1)r^{n_1-2} &  n_2(n_2-1)r^{n_2-2}& n_3(n_3-1)r^{n_3-2} &  
%n_4(n_4-1)r^{n_4-2}\\
%n_1(n_1-1)(n_1-2)r^{n_1-3} & n_2(n_2-1)(n_2-2)r^{n_2-3} &  n_3(n_3-1)(n_3-2)r^%{n_3-3}& n_4(n_4-1)(n_4-2)r^{n_4-3}
%\end{pmatrix}
%\end{equation}
%
The exponents $n_i \in \mathbb{C}$ are determined by the fourth-order 
polynomial obtained by inserting $\mu(r)=r^{n_i}$ into the left-hand 
side of (\ref{eq:bzero}) and equating this to zero. 
The solutions are 
\begin{alignat}{2}
& n_1 &&\,=\, 7 \\
& n_2 &&\,=\, 2 - \frac{11}
  {{\left( -72 + {\sqrt{6515}} \right) }^{\frac{1}{3}}} 
 +{\left( -72 + {\sqrt{6515}} \right) }^{\frac{1}{3}} \\
& n_3 &&\,=\, 2 + \frac{11\,\left( 1 + i \,{\sqrt{3}} \right)}%
 {2\,{\left( -72 + {\sqrt{6515}} \right) }^{\frac{1}{3}}} 
 - \frac{\left( 1 - i \,{\sqrt{3}} \right)\,
  {\left( -72 + {\sqrt{6515}} \right) }^{\frac{1}{3}}}{2} \\
& n_4 &&\,=\, 2 + \frac{11\,\left( 1 - i \,{\sqrt{3}} \right)}%
 {2\,{\left( -72 + {\sqrt{6515}} \right) }^{\frac{1}{3}}} 
 - \frac{\left( 1 + i \,{\sqrt{3}} \right)\,
   {\left( -72 + {\sqrt{6515}} \right) }^{\frac{1}{3}}}{2} 
\end{alignat}
Note that $n_1,n_2$ are real and $n_3,n_4$ are complex conjugates.
Of course we could construct two real solutions from the
complex ones but, as already argued, we may continue to use a 
complex $\phi$. This is advisable since the complex form (\ref{fuma}),
whose entries do not contain sums, has a simpler inverse. 
The fourth column of the inverse is the vector $\vec v$. It reads:
\begin{equation}
\label{eq:VecV}
\vec v(r)=
\begin{pmatrix}
\frac{r^{3-n_1}}{(n_1-n_3)(n_1-n_4)(n_1-n_2)} \\
\frac{r^{3-n_2}}{(n_2-n_3)(n_2-n_4)(n_2-n_1)} \\
\frac{r^{3-n_3}}{(n_3-n_4)(n_3-n_1)(n_3-n_2)} \\
\frac{r^{3-n_4}}{(n_4-n_3)(n_4-n_1)(n_4-n_2)}
\end{pmatrix}
\end{equation}

We can now explicitly evaluate the integral condition (\ref{phic}).
This leads to 
\begin{equation}
\label{eq:IntCond}
\begin{split}
\int_1^2 dr\
r^{3-n_i}\Bigl(
&-\,\frac{1}{2}r^2\alpha^{(6)}(r)+r\alpha^{(5)}(r)+4\alpha^{(4)}(r)\\
&-\,\frac{28}{r}\alpha^{(3)}(r)+\frac{72}{r^2}\alpha^{(2)}(r)
-\frac{72}{r^3}\alpha^{(\prime)}(r)\Bigr)=0\\
\end{split}
\end{equation}
which is to be read as a condition for $\alpha$, where the $n_i$ take the 
values computed above. Performing the partial integrations 
explicitly we can state the conditions in the form 
\begin{equation}
\label{foco}
\int_1^2 dr\ w_i(r) \alpha^\prime(r)=0
\qquad\text{or}\quad
\int_1^2 w_i^\prime \alpha(r)=-w_i(1)\,,
\end{equation}
where the $w_i$ are four real, non constant, linearly independent 
functions on $[1,2]$, whose exact form need not be spelled out here.  
Regarding our functions as elements of the Hilbert space 
$L^2([1,2],dr)$ this states that we need to choose $\alpha$ with 
prescribed orthogonal projections onto the four $w_i$. This can 
always be achieved due to their linear independence (for linear 
dependent $w_i$ the projection conditions might turn out to be 
contradictory), while still leaving an infinite freedom for 
$\alpha$. Hence we have shown that for $b=0$ the functions 
$\alpha$ and $\mu$ can be related such that the DN-problem is 
solved at first order. This we summarize as 

\vspace{0.3cm}
\noindent
\textbf{Lemma:} The DN-problem is solvable at the first order in the
inverse glueing radius for the choice $b=0$. In this case the glued 
and original data have the same overall ADM-energy.
\par\vspace{0.3cm}

\subsection{The case $b>0$}
\label{sec:CasebPos}
We now wish to extend the result to small positive $b$,
which corresponds to a reduction of energy. The idea is 
to set up an implicit-function theorem. For this it is 
convenient to explicitly display the relevant dependencies 
on the parameter $b$, which we suppressed so far.  

Pick a glueing function $\alpha_0(r)$ which satisfies the four
integral conditions (\ref{foco}). Define the Banach space of
bump-functions
\begin{equation}
\label{eq:BanachBump} 
\mathcal{B}:=\bigl\{\beta\in C^\infty([1,2])\ \mid \
\beta^{(n)}(1)=\beta^{(n)}(2)=0\quad\forall n\in\mathbb{N}_0\bigr\}\,,
\end{equation}
equipped with the $max$-norm on $[1,2]$. Next consider a map
$f:\mathbb{R}\times\mathcal{B}\rightarrow\mathbb{R}^4$, defined by: 
\begin{equation} 
\label{bama}
f\,:\,(b,\beta)\mapsto \Phi(b,r=2)\int_1^2dt\
\vec v(b,t)\,\mathcal{F}[b,\alpha_0+\beta](t)\,.
\end{equation}
As already argued, this map indeed maps into $\mathbb{R}^4$ even 
if we use complex fundamental matrices. Note also that the function 
$\alpha_0(t)+\beta(t)$ is a gluing function. The map (\ref{bama}) 
is a smooth map of Banach spaces, which is linear in 
$\alpha_0+\beta$. 

We have shown above that $f(0,0)=0$. We wish to show that for any small 
$\tilde{b}\in\mathbb{R}$ there exists a $\tilde{\beta}\in\mathcal{B}$ 
such that $f(\tilde{b},\tilde{\beta})=0$. This is ensured if the 
differential $D_2f(0,0)\,:\,\mathcal{B}\rightarrow \mathbb{R}^4$ is 
surjective. Since the map $f$ is linear in $\alpha_0+\beta$, we arrive at
\begin{equation} 
\label{linbama}
D_2f(0,0)\,:\,\tilde{\beta}\mapsto\Phi(0,r=2)\int_1^2 dr\ 
\vec v(r)\,\mathcal{F}[0,\tilde{\beta}](r)\,,
\end{equation}
where $\vec v(0,r)=\vec v(r)$ is explicitly given by the expression 
in (\ref{eq:VecV}). This indeed shows surjectivity and we have 

\vspace{0.3cm}
\noindent
\textbf{Theorem:} For sufficiently small $b \in \mathbb{R}$ 
the Dirichlet-Neumann-problem can be solved to first order 
in the inverse gluing radius. Hence the gluing can be used to 
reduce the ADM-energy.
\par\vspace{0.3cm}

\section{Conclusion}
\label{sec:Conclusion}
Motivated by recent results concerning the radiation content at 
spacelike infinity of the most simple, time symmetric, conformally 
flat two-black-hole data (Brill-Lindquist data), we considered 
modifications of that data which render then Schwarzschild at 
infinity by using Corvino's gluing technique. We used Brill waves 
for the gluing and have given a perturbative argument (for large 
gluing radii) that this technique may be used to reduce the ADM-energy. 
We consider this as being a first small step towards making Corvino's 
technique more concrete, which also sheds some light onto the 
problem of how to remove `spurious' radiation at spacelike infinity. 
A natural extension of this work would obviously consist in giving 
a rigorous proof for the possibility of energy reduction beyond 
finite-order approximation techniques.   

More generally, an interesting question to ask is what happens if 
one goes to smaller and smaller gluing radii. Could one reasonably 
expect this procedure to lead to a minimal overall ADM-energy for 
given hole-masses and -separations, so as to correspond to the 
ideal situation of no `spurious' radiation? Or should one rather 
expect such a minimum not to exist? We do not know the answer to 
this question.

\end{document}